%% file: delay_arxiv_0225.tex
\documentclass[journal]{IEEEtran}
\usepackage{amsmath}
\usepackage{amsfonts}
\usepackage{latexsym}
\usepackage{subfigure}
\usepackage{mathrsfs}
\include{epsf}

\newtheorem{theorem}{Theorem}[section]
\newtheorem{lemma}{Lemma}[section]
\newtheorem{corollary}{Corollary}[section]
\newtheorem{definition}{Definition}[section]

\usepackage[dvips]{graphicx}

\begin{document}

\title{Throughput-Delay Trade-off for Hierarchical Cooperation in Ad Hoc Wireless Networks}
\author{Ayfer {\"O}zg{\"u}r, Olivier L\'ev\^eque,~\IEEEmembership{Member,~IEEE}
\thanks{Manuscript received ??; revised ??, and ??.
The work of Ayfer {\"O}zg{\"u}r was supported by Swiss NSF grant Nr 200020-118076. The material in this paper was presented in part at the IEEE International Conference on Telecommunications, Saint-Petersburg, June 2008.}
\thanks{The authors are with the Ecole Polytechnique F\'ed\'erale de Lausanne, Facult\'e Informatique et Communications, Building INR, Station 14, CH - 1015 Lausanne, Switzerland (e-mails: \{ayfer.ozgur,olivier.leveque\}@epfl.ch).}
}

\maketitle

\begin{abstract}
Hierarchical cooperation has recently been shown to achieve better throughput scaling than classical multihop schemes under certain assumptions on the channel model in static wireless networks. However, the end-to-end delay of this scheme turns out to be significantly larger than those of multihop schemes. A modification of the scheme is proposed here that achieves a throughput-delay trade-off $D(n)=(\log n)^2 T(n)$ for $T(n)$ between $\Theta(\sqrt{n}/\log n)$ and $\Theta(n/\log n)$, where $D(n)$ and $T(n)$ are respectively the average delay per bit and the aggregate throughput in a network of $n$ nodes. This trade-off complements the previous results of El Gamal et al.,~which show that the throughput-delay trade-off for multihop schemes is given by $D(n)=T(n)$ where $T(n)$ lies between $\Theta(1)$ and $\Theta(\sqrt{n})$.

Meanwhile, the present paper considers the network multiple-access problem,  which may be of interest in its own right.

\end{abstract}


\section{Introduction}\label{sec:intro}

Scaling laws offer a way of studying fundamental trade-offs in wireless networks as well as of highlighting the qualitative and architectural properties of specific designs. Such study has been initiated by the work \cite{GK00} of Gupta and Kumar in 2000. Their by now familiar model considers $n$ nodes randomly distributed on a unit area, each of which wants to communicate to a random destination at a common rate $R(n)$. They ask what is the maximally achievable scaling of the aggregate throughput $T(n)=nR(n)$ and show that cooperation between nodes can dramatically improve performance. Instead of using the simple scheme of time-sharing between direct transmissions from source nodes to destinations, which only achieves aggregate throughput $\Theta(1)$, the nodes can cooperate and relay the packets by multihopping from one node to the next, in which case an aggregate throughput scaling of $\Theta(\sqrt{n})$ is achieved. The price to pay, however, is in terms of delay. In the multi-hop scheme, the packets need to be retransmitted many times before they reach their actual destinations, which results in larger end-to-end delay. More precisely, as shown later in \cite{GMPS06-I,GMPS06-II}, in a multi-hop scheme, bits are delivered to their destinations in $\Theta(\sqrt{n})$ time-slots on average after they leave their source nodes, while the average delay for the simple TDMA scheme remains only $\Theta(1)$. Note that this accounts only for on-the-flight delay; the queuing delay at the source node is not considered.

Recently, it has been shown in \cite{OLT07} that under certain assumptions on the channel model, a much better throughput scaling is achievable in wireless networks than the one achieved by classical multi-hop schemes. The authors exhibit a hierarchical cooperation scheme that uses distributed MIMO communication to achieve aggregate throughput scaling arbitrarily close to linear, i.e. $T_h(n)=\Theta(n^{\frac{h}{h+1}})$ for any integer $h>0$. The parameter $h$ corresponds to the number of hierarchical levels used in the scheme and by increasing $h$, one can get arbitrarily close to linear scaling. A natural question is whether there is a price to pay for this superior scaling of the throughput. In particular, what is the throughput-delay trade-off for hierarchical cooperation? In this paper, we analyze the delay performance of the hierarchical cooperation scheme and show that the structure suggested in \cite{OLT07} is very suboptimal from the delay point of view. We propose a modification of the scheme in this paper, that achieves much better delay performance for the same throughput. More precisely, we show that one important drawback of the scheme in \cite{OLT07}, that is not immediately clear from the presentation in there, is that it uses an extremely large bulk-size, where the bulk-size of a scheme refers to the minimum number of bits that should be communicated between each source-destination pair under this scheme. The bulk-size used by the scheme in \cite{OLT07} scales as $B_h(n)=\Theta(n^{\frac{h}{2}})$; in other words, it grows arbitrarily fast as the throughput approaches linear scaling. Note that the bulk-size immediately imposes a lower bound on the end-to-end delay of each communication; even if there is no transmission delay from the source node to the destination node, receiving a bulk of $B(n)$ bits will take at least $\Theta(B(n)/\log n)$ channel uses for a destination node, since a simple application of the cut-set bound upper bounds the rate of reception by (or transmission from) a node with $\log n$ bits per channel use.

The basic idea behind the hierarchical cooperation scheme in \cite{OLT07} is to first distribute the bits of a source node among its neighboring nodes, so that these bits can then be simultaneously transmitted to a group of nodes in the vicinity of the destination node. By collecting the observations of the receiving nodes to the actual destination node, the destination node is able to recover the bits intended for itself. This kind of transmission is often referred to as distributed MIMO since it resembles the multiple-input-multiple-output transmissions between a transmitter and receiver pair with multiple transmit and receive antennas respectively. The efficiency of the distributed MIMO transmission increases with the size of (number of nodes contained in) the transmit and receive clusters, formed around the source node and the destination node respectively. However, if the size of the transmit cluster is large, the bulk of data to be communicated by each source node has to be large enough to be chopped off and distributed among the many nodes in the cluster. Hence, the size of the transmit cluster imposes a lower bound on the bulk size that needs to be communicated between each source-destination pair. Moreover, distributing the bits of the source node before the MIMO transmission and collecting the observations to the destination node following the MIMO transmission brings another traffic requirement. It has been shown in \cite{OLT07} that this cooperation traffic can be handled efficiently if decomposed into multiple problems of the original kind, i.e., of communicating between $n$ source-destination pairs in a network of $n$ nodes and reusing the idea of distributed MIMO. This recursion builds a hierarchical architecture that is shown to be efficient from throughput point of view. However, since distributed MIMO based communication imposes a lower bound on the bulk-size, repeating the idea recursively yields a scheme with even larger bulk-size. This is the reason why the bulk size of the hierarchical cooperation scheme increases as $\Theta(n^{\frac{h}{2}})$ with $h$ hierarchical levels.

In this paper, we suggest a modification of the hierarchical cooperation scheme in \cite{OLT07} that handles the problem of cooperation more efficiently. In order to do this, we study the problem of cooperation more carefully by posing it as a network multiple access problem, instead of separating it into multiple unicast problems as was originally done in \cite{OLT07}. In the network multiple access problem, each of the $n$ nodes in the network is interested in conveying independent information, say $L$ bits, to each of the other nodes in the network. We propose a two-phase hierarchical scheme that solves the multiple access problem in $\Theta(n^{\frac{h+1}{h}})$ time-slots for any $h>0$. Using this scheme for cooperation, the modified hierarchical cooperation scheme achieves the same aggregate throughput $T_h(n)=\Theta(n^{\frac{h}{h+1}})$ by using a much smaller  bulk-size $B_h(n)=T_h(n)$. We show that reduced bulk size consequently reduces the delay and the modified hierarchical cooperation scheme achieves $D_h(n)=\Theta(n)$.

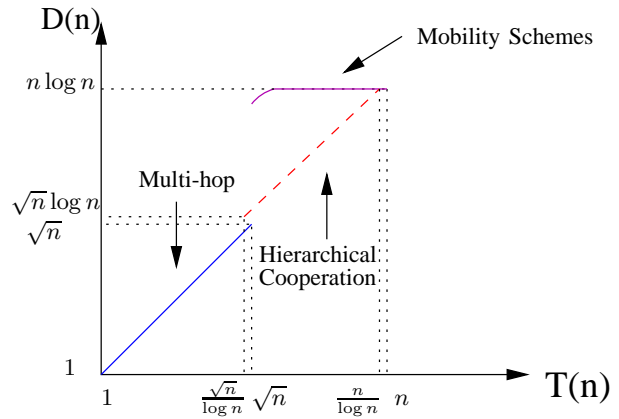
\begin{figure}
\begin{center}
\input{delay-thrput_tradeoff_3.pstex_t}
\end{center}
\caption{Throughput-delay performance achieved by hierarchical cooperation together with known results from the literature.}
\label{fig:delay2}
\end{figure}

We proceed by optimizing scheduling in this scheme to further reduce the end-to-end delay. To do this, we need to consider a generalized version of the multiple access problem where each node in the network is interested in conveying independent information to each of the nodes in a subset of $A(n)$ nodes, where the $A(n) < n$ nodes are chosen uniformly at random among the $n$ nodes in the network. We show that this task can be accomplished in $\Theta(\frac{A(n)}{n}n^{\frac{h}{h+1}}\log n)$ channel uses for any $h>0$ if $A(n)\geq n^{\frac{h}{h+1}}$. This allows us to achieve a throughput delay trade-off of $(T(n),D(n))=(n^b/\log n,n^b\log n)$ for any $0\leq b< 1$. This new result is depicted in Figure~\ref{fig:delay2}, together with previous results from the literature. 

A related line of research (see e.g. \cite{GMPS06-I, GT02, NM05,  SMS06}) is the characterization of the throughput-delay trade-off for mobile networks, where nodes move over the duration of communication according to a certain mobility pattern. In general, mobility schemes achieve an aggregate throughput scaling comparable to that of hierarchical cooperation (i.e. up to linear in $n$), but the delay scaling performance of such schemes may vary significantly, depending on the chosen mobility model. For instance, under the classical random walk mobility model considered in \cite{GMPS06-I}, the performance is quite poor, as illustrated in Figure~\ref{fig:delay2}. But from the delay point of view, a more prominent disadvantage which is common to all mobility models and which does not appear on the graph in Figure~\ref{fig:delay2}, is the constant that precedes the delay scaling law. Roughly speaking, this pre-constant relates to the speed of nodes in the case of mobility schemes, whereas it relates to the speed of light in the case of hierarchical cooperation.

\section{Setting and Main Results} \label{sec:result}

There are $n$ nodes uniformly and independently distributed in a square of unit area. Every node is both a source and a destination. The sources and destinations are paired up one-to-one in a random fashion without any consideration on respective locations. Each source has the same traffic rate $R(n)$ to send to its destination node. (In the following text, we will sometimes refer to this traffic pattern as the unicast problem in order to distinguish it from the multicast problems that will be discussed in Sections~\ref{sec:sbs} and \ref{sec:bshcs}.) The aggregate throughput of the system is $T(n) = n R(n)$.

The complex baseband-equivalent channel gain between  node $i$ and node $k$ at time $m$ is given by:
\begin{equation}
\label{eq:ch_model} H_{ik}[m] = r_{ik}^{-\alpha/2} \exp(j
\theta_{ik}[m])
\end{equation}
where $r_{ik}$ is the distance between the nodes, $\theta_{ik}[m]$ is the random phase at time $m$, uniformly distributed in $[0,2\pi]$ and $\{\theta_{ik}[m], 1\leq i\leq n, 1\leq k\leq n\}$ is a collection of i.i.d. random processes. The $\theta_{ik}[m]$'s and the $r_{ik}$'s are also assumed to be independent. The constant $\alpha \ge 2$ is called the power path loss exponent of the environment. 

Note that the channel is random, depending on the location of the users and the phases. The locations are assumed to be fixed over the duration of the communication. The phases are assumed to vary in a stationary ergodic manner (fast fading). We assume that the channel gains are known at all the nodes. The signal received by node $i$ at time $m$ is given by
$$
Y_i[m]=\sum_{k=1}^{n}H_{ik}[m]X_k[m]+Z_i[m]
$$
where $X_k[m]$ is the signal sent by node $k$ at time $m$ and $Z_i[m]$ is white circularly symmetric Gaussian noise of variance $N_0$ per symbol. Every node is subject to a transmit power constraint that we denote by $P$.\footnote{We present the low-level assumptions on the channel and network model in this section for the sake of completeness. However, most of the discussions in the following sections will rely on intermediate results established in \cite{OLT07}, hence the dependence of the results on the low level assumptions might not be always clear.} 

\bigbreak

The delay $D(n)$ of a communication scheme for this network is defined as the average time it takes for a bit, or packet of constant size, to reach its destination node after it leaves its source node, where the average is taken over all bits or packets traveling in the network. So defined, the delay of a scheme quantifies the average time spent by the bits traveling inside the network while operated under this scheme.

This definition of delay is consistent with \cite{GMPS06-I,GMPS06-II} and therefore the comparison in Figure~\ref{fig:delay2} of the multihop scheme and hierarchical cooperation is fair. However, note that this definition does not include the queuing delay at the source node, as the clock starts when a packet leaves its source node. The delay at the source node can be accounted for by assuming a particular packet arrival process and studying the overall delay of a packet from its arrival at the source queue to the decoding at the destination node. The transmission delays given in Figure~\ref{fig:delay2} can be regarded as lower bounds to this overall delay. Consider for example the simple TDMA scheme, one at a time transmission between the source-destination pairs, that corresponds to the origin in Figure~\ref{fig:delay2}. Assume independent Poisson packet arrival at each source node of appropriate rate. If we assume round-robin fashion, backlog unaware scheduling between the transmissions, the overall delay of the TDMA scheme will be $\Theta(n)$ much larger than the $\Theta(1)$ delay predicted by Figure~\ref{fig:delay2}. However, it is known that this delay can be reduced to $O(\log n)$  with backlog aware scheduling \cite{NMC07}. In general, how larger is the overall delay from the transmission delay given in Figure~\ref{fig:delay2} depends on how well we can match the packet arrival process with backlog aware scheduling schemes. In this paper, our aim is to quantify the transmission delay of the discussed schemes; the second question regarding the queuing delay at the source is left open.

\bigbreak
The following theorem is the main result of this paper.

\begin{theorem} \label{thm:dense}
Using hierarchical cooperation, the following points are achievable on the throughput-delay scaling curve,
$$
(T(n),D(n)) = \Theta \left( n^b/\log n, n^b\log n\right)
$$
where $0\leq b<1$ (see Figure \ref{fig:delay2}).
\end{theorem}

\section{Overview of the Hierarchical Cooperation Scheme}\label{sec:hier}

In this section, we give a brief overview of the hierarchical cooperation scheme as presented in \cite{OLT07} and establish the throughput-delay trade-off for this scheme. Some of the discussions presented here directly build on results already established in \cite{OLT07}.
\bigbreak

The hierarchical cooperation scheme is based on clustering the nodes in the network and performing long-range MIMO transmissions between the clusters. The long-range MIMO transmissions should be proceeded and followed by cooperation phases establishing transmit and receive cooperation respectively, which yields three successive phases in the operation of the network. If simple TDMA is used for establishing cooperation in phase 1 and 3, the overall scheme achieves a $\sqrt{n}$-scaling of the aggregate throughput. This is the three phase scheme discussed in Section~\ref{sec:hier1}. Higher throughputs can be achieved by setting the cooperation problem as multiple communication problems and using the three phase scheme as a solution to each of those communication problems. This yields the idea of recursion and results in a hierarchical architecture, where increasing the number of levels in the hierarchy yields an aggregate throughput scaling arbitrarily close to linear. The hierarchical cooperation scheme is discussed in more detail in Section~\ref{sec:hier2}. 

\subsection{The Three Phase Scheme}\label{sec:hier1}
The network is divided into clusters of $M_1$ nodes and a particular source node $s$ sends $M_1$ bits to its destination node $d$ in three steps:

\begin{itemize}
\item [(1)] Node $s$ first distributes its $M_1$ bits among the $M_1$ nodes in its cluster, one bit for each node;

\item [(2)] These nodes together can then form a distributed transmit antenna array, sending the $M_1$ bits {\em simultaneously} to the destination cluster where $d$ lies;

\item [(3)] Each node in the destination cluster gets one observation from the MIMO transmission, and it quantizes and ships the observation to $d$, which can then do joint MIMO processing of all the observations and decode the $M_1$ transmitted bits.
\end{itemize}

From the network point of view, all source-destination pairs have to
eventually accomplish these three steps. Step 2 is long-range
communication and only one source-destination pair can operate at
a time. Steps 1 and 3 involve local communication and can be
parallelized across source-destination pairs. Combining all this
leads to the following three phases in the operation of the network:

\textbf{Phase 1: Setting Up Transmit Cooperation} Clusters work in
parallel. Within a cluster, each source node distributes $M_1$
bits to the other nodes, $1$ bit for each node, such that at the end
of the phase, each node has $1$ bit from each of the other nodes in
its cluster. (Recall our assumption that each node is a source
for some communication request and a destination for another.) Thus, since
there are $M_1$ source nodes in each cluster, this gives a traffic demand
of exchanging $M_1(M_1-1)\sim M_1^2$ bits. Using TDMA, one-at-a-time transmission between
pairs of nodes, these $M_1^2$ bits can be exchanged in $M_1^2$ time slots.\footnote{Note that although because of the broadcasting nature of TDMA, every bit of a source node can be conveyed to all other nodes in the cluster for free, this is not what we require here. In the MIMO transmissions that are following in the next phase, every node is independently encoding its data and it does not need to know the bits transmitted by the other nodes. A standard reference on the capacity of MIMO channels is \cite{Tel99}. The derivations for the current case can be found in \cite{OLT07}.}

\textbf{Phase 2: MIMO Transmissions}
Successive long-distance MIMO transmissions are performed between
source-destination pairs, one at a time. In each one of the MIMO transmissions,
say the one between $s$ and $d$, the $M_1$ bits of $s$ are simultaneously
transmitted by the $M_1$ nodes in its cluster to the $M_1$ nodes in the
cluster of $d$. Each of the long-distance MIMO transmissions are
repeated for each source-destination pair in the network, hence
we need $n$ time-slots to complete the phase.

\textbf{Phase 3: Cooperate to Decode} Clusters work in parallel.
Since there are $M_1$ destination nodes inside the clusters, each
cluster received $M_1$ MIMO transmissions in phase 2, one intended for
each of the destination nodes in the cluster. Thus, each node in the
cluster has $M_1$ received observations, one from each of the MIMO
transmissions, and each observation is to be conveyed to a different
node in its cluster. Nodes quantize each observation into
fixed $Q$ bits, so there are a total of $QM_1^2$ bits to be exchanged
inside each cluster. Using TDMA as in Phase 1, the phase can be
completed in $QM_1^{2}$ time slots.\footnote{In order to be able to convey the salient features of the hierarchical cooperation scheme to the reader in the simplest way, a rather informal approach is taken in this section and some technical details are omitted. A rigorous description of the scheme can be found in \cite{OLT07}. For example, it is not necessary to have exactly $M_1$ nodes in each cluster but it suffices to have $\Theta(M_1)$ nodes for a scaling law analysis. It is shown in \cite{OLT07} that by dividing the network into cells of certain area, we can ensure having $\Theta(M_1)$ nodes in all cells with high probability. Moreover, the case when a source node and its destination lie in the same cluster should be treated separately. Similarly, assuming that each source node is sending exactly $1$ bit to each of the other nodes in its cluster in phase 1 is a simplification. A rigorous argument will assume that each source node is sending $L$ bits to each of the other nodes where $L$ is a large enough constant independent of $M_1$ and $n$. The rates of the TDMA transmissions in phase 1 and phase 3 and the per node rate for the MIMO transmission in phase 2 are assumed to be $1$ for simplicity, so that $1$ bit is transmitted in $1$ time slot. The actual rates of these transmissions can be shown to be constants depending on the system parameters and independent of $M_1$ and $n$. Also, in phase 1 and phase 3 not all clusters should be allowed to operate simultaneously but a TDMA scheme between the clusters should be employed so that the resultant inter-cluster interference is bounded and each cluster becomes active a constant fraction of time.}

In \cite{OLT07}, it is shown that each destination node is able to decode the
transmitted bits from its source node from the $M_1$ quantized signals
it gathers by the end of Phase 3. The throughput achieved by the scheme can be calculated as follows: each source node is able to transmit $M_1$ bits to its destination node, hence $nM_1$ bits in total are delivered to
their destinations in $M_1^{2}+n+QM_1^{2}$ time slots, yielding an
aggregate throughput of $$ \frac{nM_1}{M_1^{2}+n+ QM_1^{2}}
$$ bits per time-slot. Choosing
$M_1=\sqrt{n}$ to maximize this expression yields an aggregate throughput $T(n)=\frac{1}{2+Q}\sqrt{n}$.

Note that as opposed to multihop, this three phase scheme allows only bulk transmission between any source-destination pair in the network; i.e. one can not arbitrarily communicate one bit (or $L$ bits with $L$ constant) using the three-phase scheme, but $M_1=\sqrt{n}$ bits should be transferred between every source-destination pair  with each use of the scheme. 

The end-to-end delay of this scheme is simply the total time for the three phases, since the bits are leaving their source nodes at the beginning of the first phase and are only decoded by their respective destination nodes at the end of the third phase. With the choice $M_1=\sqrt{n}$, we see that the delay of the three phase scheme is $D(n)=(2+Q)n$. Note that this delay scaling is much worse when compared to the delay of the multi-hop scheme achieving same aggregate throughput. 

\subsection{The Hierarchical Cooperation Scheme}\label{sec:hier2}

Higher aggregate throughput scaling can be achieved by using better network communication schemes than TDMA to establish the transmit and receive cooperations in the first and third phases of the three phase scheme described in the previous section. Recall that there are $M_1^2$ and $QM_1^2$ bits to be exchanged inside each cluster in phases 1 and 3, respectively. This traffic demand of exchanging $M_1^2$ bits (or $QM_1^2$ bits) can be handled by setting up $M_1$ sub-phases, and assigning $M_1$ pairs in each sub-phase to communicate their $1$ bit (or $Q$ bits). The traffic to be handled at each sub-phase now looks similar to the original network communication problem (the unicast network problem defined in Section~\ref{sec:result}), with $M_1$ users instead of $n$. Any scheme suggesting a good solution for the original problem can now be used inside the sub-phases as an alternative to TDMA; for example, the multi-hop scheme and the three-phase scheme itself would be two alternatives both achieving an aggregate throughput scaling $\Theta(\sqrt{M_1})$ (in a network of size $M_1$) as opposed to the $\Theta(1)$ scaling achieved by TDMA. 

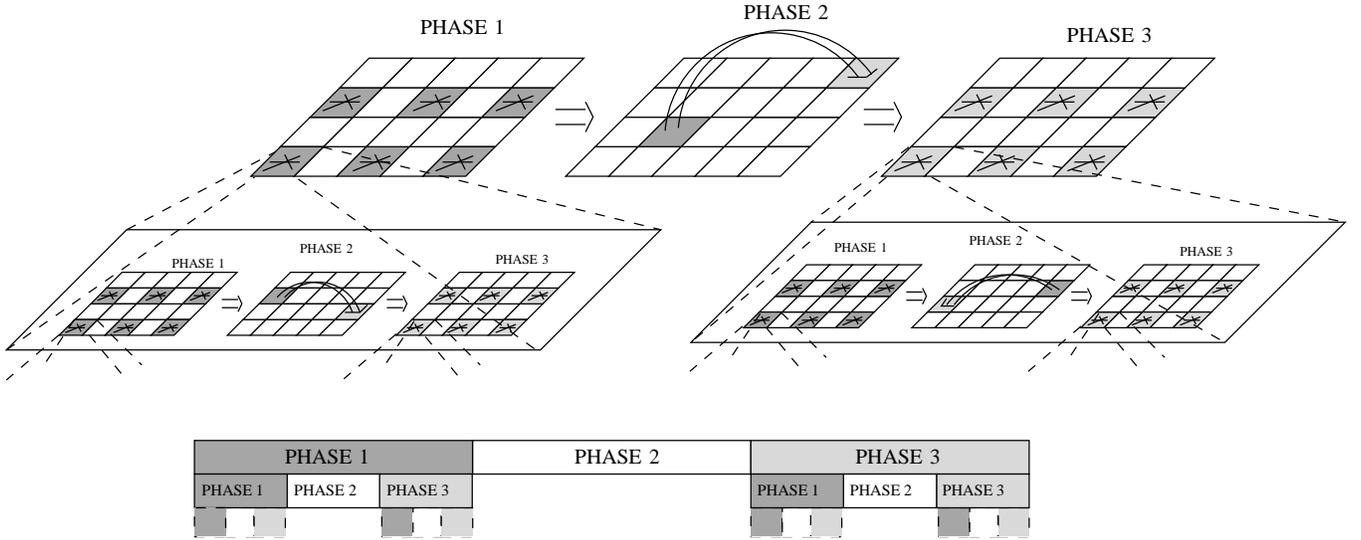
\begin{figure*}[tb]
\begin{center}
\input{phase1_3d_ver2.pstex_t}
\end{center}
\caption{The salient features of the three phases and the time division in a hierarchical scheme are illustrated. Figure taken from \cite{OLT07}.}
\label{fig:hier3d}
\end{figure*}

Consider using the three phase scheme for cooperation as suggested in \cite{OLT07}. More precisely, we want to handle the traffic of communicating $1$ bit (or $Q$ bits) between the $M_1$ pairs assigned in each sub-phase of phase 1 (or phase 3), by further dividing the clusters into smaller clusters of size $M_2$ and reusing the three phase scheme (TDMA-MIMO-TDMA). Note that this will create a hierarchical structure with two levels. See Figure~\ref{fig:hier3d}. Note however that the three phase scheme in Section~\ref{sec:hier1}, allows only bulk transmissions between source-destination pairs. In this particular case, one will have to communicate $M_2$ bits between the source-destination pairs assigned at each sub-phase, as opposed to the original requirement of communicating only $1$ bit (or $Q$ bits). For the overall scheme, this in turn increases the bulk size to be communicated between every source-destination pair in the network from $M_1$ bits to $M_1\times M_2$ bits. So for the 2-level hierarchical scheme, we have to start by assuming that each source node in the network has $M_1\times M_2$ bits to communicate to its destination node. It can be seen that these $M_1\times M_2$ bits per source destination pair, or a total $n\times M_1\times M_2$ bits in the network, can be communicated in  
\begin{equation}\label{time-2h}
M_1(M_2^2+M_1+QM_2^2)+M_2n+M_1Q(M_2^2+M_1+QM_2^2)
\end{equation}
time slots using the 2-level hierarchical scheme. The first term $M_1(M_2^2+M_1+QM_2^2)$ is the completion time of phase-1 of the three phase scheme. It is divided into $M_1$ sessions; in each session, $M_1$ source-destination pairs are assigned to communicate their $M_2$ bits using a three phase scheme of TDMA-MIMO-TDMA. Recall from the computations of the three phase scheme in Section~\ref{sec:result} that this takes $M_2^2+M_1+QM_2^2$ time slots ($M_1$ and $M_2$ here correspond to the $n$ and $M_1$, respectively, of the previous section). A similar argument holds for the third term $M_1Q(M_2^2+M_1+QM_2^2)$ in (\ref{time-2h}) which is the completion time for phase-3 with the extra $Q$ factor. Note that at the end of the first phase, each source node has distributed its $M_1\times M_2$ bits among the $M_1$ nodes in its cluster, hence $M_2$ bits for each node. These bits can be relayed to the destination cluster in $M_2$ successive MIMO transmissions. Since the MIMO transmissions have to be repeated for each of the $n$ source-destination pairs in the network, the completion time of the second phase is $M_2n$ in (\ref{time-2h}). 

Therefore, the aggregate throughput of the 2-level scheme is given by the expression
\begin{equation}\label{eq:thr_2l}
\frac{M_2\, M_1\,n}{M_1(M_2^2+M_1+QM_2^2)+M_2n+M_1Q(M_2^2+M_1+QM_2^2)} 
\end{equation}
and the optimal choices of $M_1=n^{2/3}$ and $M_2=n^{1/3}$ maximize the aggregate throughput scaling to 
$$T_2(n)=M_1=n^{2/3},$$
while the denominator dictating the delay of the scheme is of order $$D_2(n)=M_2\times n=n^{4/3}.$$ Note that with the 2-level hierarchical scheme, we improve the aggregate throughput scaling from $\sqrt{n}$ for the three-phase scheme in the previous section to $n^{2/3}$, at the cost of increasing the bulksize from $\sqrt{n}$ to $n$, which, in turn, increases the delay from $n$ to $n^{4/3}$.

The argument can be applied recursively to build an $h$-level hierarchical scheme. The optimal cluster size at the $k$'th level of an $h$-level hierarchical scheme can be computed as $M_k=n^{\frac{h+1-k}{h+1}}$. The aggregate throughput achieved by an $h$-level hierarchical cooperation scheme is given by $$T_h(n)=M_1=n^{\frac{h}{h+1}}.$$ The bulk-size is $$B_h(n) = M_h \times \ldots \times M_1 = n^{\frac{h}{2}}$$ and the end-to-end delay is $$D_h(n)=M_h\times M_{h-1}\times\dots\times M_2\times n=n^{\frac{h^2+h+2}{2(h+1)}}$$
where we observe that for large $h$, the delay exponent is linear in $h$. 

\bigbreak

\medbreak
The results obtained in this section establish the poor delay performance of hierarchical cooperation. Note that the delay is mostly due to the large bulk-size used by the scheme. This is different from multi-hop schemes since their bulk-size is constant ($\Theta(1)$) independent of the throughput achieved. The delay in multihop is rather due to the time spent in relaying the messages inside the network. In the next section, we suggest a modification of the scheme so that it achieves the same throughput using much smaller bulk-size.

\section{Hierarchical Cooperation with Smaller Bulk-Size} \label{sec:sbs}

In this section, we treat the problem of cooperation in the three phase scheme with more care. We start by defining the network multiple access problem to be the following. 

\smallbreak
\begin{definition}[The Network Multiple Access Problem]\label{def:nma}
Consider the assumptions on the network and channel model given in Section~\ref{sec:result}. Let each node in the network be interested in communicating independent information to each of the other nodes in the network. In particular, let us assume that each node has an independent $1$ bit message (or $L$ independent bits, with $L$ constant) to send to each of the other nodes in the network and the quantity of interest is the smallest time $F(n)$ required to accomplish this task. This problem we refer to be the network multiple access problem. 
\end{definition}
\smallbreak
The following theorem provides an achievable solution to this problem.
\smallbreak
\begin{theorem}\label{thm1}
For any integer $h> 0$, the network MAC problem can be solved in
$$
F(n)\leq K \, n^{\frac{h+1}{h}}
$$
time-slots w.h.p.\footnote{with high probability}, for some constant $K>0$ independent of $n$.
\end{theorem}
\smallbreak

\noindent\textit{Proof of Theorem~\ref{thm1}:} Let us start by assuming that there exists a scheme that solves the multiple access problem in $F(n)=n^b$ time-slots with $b>1$. Note that one such scheme is simple TDMA that yields $b=2$. Using this existing scheme, we will construct a new scheme that yields smaller $F(n)$. 

As before, let us start by dividing the network into clusters of $M$ nodes. Let us first focus on one specific cluster $S$ and one node $d$ located outside of this cluster. In particular, all nodes in $S$ have $1$ bit to send to $d$. These bits can be communicated to $d$ in two steps:

\begin{itemize}
\item[(1)] The nodes in $S$ \emph{simultaneously} transmit their $1$ bit messages destined to $d$ forming a distributed transmit antenna array for MIMO transmission. The nodes in the destination cluster where $d$ lies, form a distributed receive antenna array for this MIMO transmission.
\item[(2)] Each node in the destination cluster obtains one observation from the MIMO transmission in the previous phase; it quantizes and ships this observation to $d$, which can do joint MIMO processing of all the observations and decode the $M$ transmitted bits from the nodes in $S$.
\end{itemize}

As a first step towards handling the whole network problem, note that these two steps should be accomplished between $S$ and all other nodes in the network. This can again be done in two steps:

\textbf{Phase 1: MIMO transmissions} We perform successive long-distance MIMO transmissions between $S$ and all other nodes in the network. In each of the MIMO transmissions, say between $S$ and $d$, the $M$ nodes in $S$ are simultaneously transmitting the $1$ bit messages they would like to communicate to $d$ and the $M$ nodes in the cluster where $d$ lies are observing the MIMO transmission. The MIMO transmissions should be repeated for each node in the network, hence we need $n$ time-slots to complete the phase. 
 
\textbf{Phase 2: Cooperate to decode} Clusters work in parallel. Since there are $M$ nodes inside each cluster, each cluster received $M$ MIMO transmissions from $S$ in the previous phase, one intended for each node in the cluster. Thus, each node in the cluster has $M$ observations, one from each of the MIMO transmissions, and each observation is intended for a different node in the cluster. Each of these observations can be quantized into $Q$ bits, with a fixed $Q$, which yields exactly the original network multiple access problem, with $M$ nodes instead of $n$. Using the scheme we assumed to exist in the beginning of the proof, this task can be completed in $QM^b$ time slots.
\medbreak

The total time we have spent during the two phases for handling the traffic originated from cluster $S$ is given by $n+QM^b$. From the network point of view, the above two steps should be completed for all $n/M$ clusters in the network. Thus, the multicasting task can be completed in $\frac{n}{M}(n+QM^b)$ time slots in total. Choosing $M=n^\frac{1}{b}$ in order to minimize this quantity yields $F(n)=(1+Q)n^{2-\frac{1}{b}}$.

\begin{figure}[tb]
\begin{center}
\includegraphics[width=0.3\textwidth]{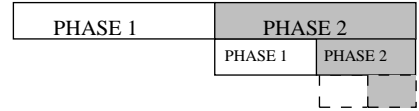}
\end{center}
\caption{The figure illustrates the time-division in the hierarchical scheme that solves the network multiple access problem.}
\label{fig:tdma_multi}
\end{figure}

Note that $2-\frac{1}{b}< b$ for $b>1$. In other words, we have established a solution for the multiple access problem that is better than the one we started with. Indeed, the two phase scheme described above can be used recursively yielding a better scheme at each step of the recursion. In particular, starting with TDMA achieving $b=2$ and applying the idea recursively $h$ times, one gets a scheme that solves the multiple access problem in $\Theta(n^\frac{h+1}{h})$ time slots. The operation of this scheme is illustrated in Figure~\ref{fig:tdma_multi}.\hfill $\square$

\begin{figure*}[tb]
\begin{center}
\includegraphics[width=0.8\textwidth]{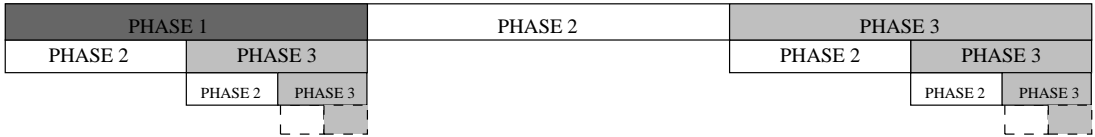}
\end{center}
\caption{The figure illustrates the time-division in the modified hierarchical scheme that uses the scheme in Figure~\ref{fig:tdma_multi} for cooperation. Note the difference in operation of the phases between the modified hierarchical cooperation scheme and the original hierarchical cooperation scheme of \cite{OLT07} in Figure~\ref{fig:hier3d}.}
\label{fig:tdma_modhier}
\end{figure*}

\medbreak
The interest in the multiple access problem arises from the fact that it exactly models the required traffic for cooperation in the three phase scheme. Recall the communication requirement inside the clusters in Phase 1 and 3 described in Section~\ref{sec:hier1}. This communication requirement, equivalent to a network multiple access problem, is handled using TDMA in the three phase scheme which has been seen to be suboptimal from throughput point of view in the Section~\ref{sec:hier1}. In the hierarchical cooperation scheme described in Section~\ref{sec:hier2}, this multiple access problem is handled by decomposing it into a number of unicast network problems. The resultant scheme is optimal in terms of throughput, but not very satisfying in terms of bulk-size. By using the solution to the multiple access problem suggested in this section, one can modify the hierarchical cooperation scheme, so as to achieve the same throughput with smaller bulk-size and consequently smaller delay. The resultant modified hierarchical scheme is illustrated in Figure~\ref{fig:tdma_modhier}. Note that the gain is coming from treating the cooperation problem as it is and not necessarily as multiple unicast problems as was previously done in Section~\ref{sec:hier2}.
\smallbreak

\begin{corollary}\label{col2}
A modified hierarchical cooperation scheme can achieve an aggregate throughput $T_h(n)\geq K_1 n^{\frac{h}{h+1}}$ with bulk-size $B_h(n)=K_2 n^{\frac{h}{h+1}}$ and delay $D_h(n)\leq K_3n$ w.h.p., for any integer $h\geq 0$ and some positive constants $K_1, K_2, K_3$ independent of $n$. 
\end{corollary}
\smallbreak
\noindent\textit{Proof of Corollary~\ref{col2}:} Consider the three phase hierarchical scheme described in Section~\ref{sec:hier1}. By Theorem~\ref{thm1}, the required traffic for transmit and receive cooperation in phase 1 and phase 3 can be handled in $KM^{\frac{h+1}{h}}$ and $KQM^{\frac{h+1}{h}}$ time slots respectively. The expression for the aggregate throughput then becomes $$\frac{Mn}{KM^{\frac{h+1}{h}}+n+KQM^{\frac{h+1}{h}}}$$ which is maximized by the choice $M=n^\frac{h}{h+1}$, yielding aggregate throughput $T_h(n)=\frac{1}{1+K+KQ}n^\frac{h}{h+1}$, bulk-size $B_h(n)=n^\frac{h}{h+1}$ and delay $D_h(n)=(1+K+KQ)n$.\hfill $\square$

\section{Hierarchical Cooperation with Better Scheduling}

In the previous section, we presented a modified hierarchical scheme that achieves throughput $T_h(n)=\Theta(n^\frac{h}{h+1})$ using bulk-size $B_h(n)=\Theta(n^\frac{h}{h+1})$. However, the delay of this scheme is still $D_h(n)=\Theta(n)$. In this section, we optimize the scheduling in the scheme to further improve the delay performance to $D_h(n)=\Theta(n^\frac{h}{h+1}\log n)$. We first start by improving the scheduling in the three phase scheme with $h=1$ discussed in Section~\ref{sec:hier1}. We then consider the modified hierarchical scheme with $h\geq 2$ discussed in Section~\ref{sec:sbs} . 

Before starting, we state the following binning lemma, similar in spirit to Lemma 4.1 and Lemma 5.1 in \cite{OLT07} and can be proven using similar techniques. The lemma will be used repeatedly throughout the rest of the paper.

\begin{lemma}\label{lem}
Let us assume that $f(n)$ balls are thrown into $n$ bins, independently and uniformly at random. The following properties are satisfied with failure probability exponentially small in $n$. 
\begin{itemize}
\item[(a)] If $\lim_{n\to\infty}\frac{f(n)}{n\log n}=\infty$, then there are $\Theta(\frac{f(n)}{n})$ nodes in each bin.
\item[(b)] If $\lim_{n\to\infty}\frac{f(n)}{n}=c$ with $c\geq 0$ a constant independent of $n$, then there are at most $O(\log n)$ nodes in each bin.
\end{itemize}
\end{lemma}

\subsection{Better Scheduling for the Three Phase Scheme}\label{sec:bstps}

Recall the operation of the three phase scheme from the point of view of a single source-destination pair $s$-$d$ as described in Section \ref{sec:hier1}: a step (1) where $s$ distributes its $M$ bits among the $M$ nodes in its cluster, followed by a step (2) where these $M$ bits are simultaneously transmitted to the destination cluster via MIMO transmission, and a step (3) where the quantized MIMO observations are collected at the destination node $d$. These three steps need to be eventually accomplished for each source-destination pair in the network. In this section, we improve the scheduling in accomplishing this task: we organize $M$ successive sessions and allow only $n/M$ source-destination pairs to complete the three steps in each session.

In the beginning of each session we randomly choose one source node from each cluster, thus $n/M$ source nodes in total. In general, the $n/M$ destination nodes corresponding to these randomly chosen source nodes can be located anywhere. However, from Lemma~\ref{lem}, we know that no more than $\log n$ of these destination nodes are located in the same cluster with high probability. We proceed by accomplishing the three steps for these chosen source-destination pairs:

\begin{figure*}[tb]
\begin{center}
\includegraphics[width=0.9\textwidth]{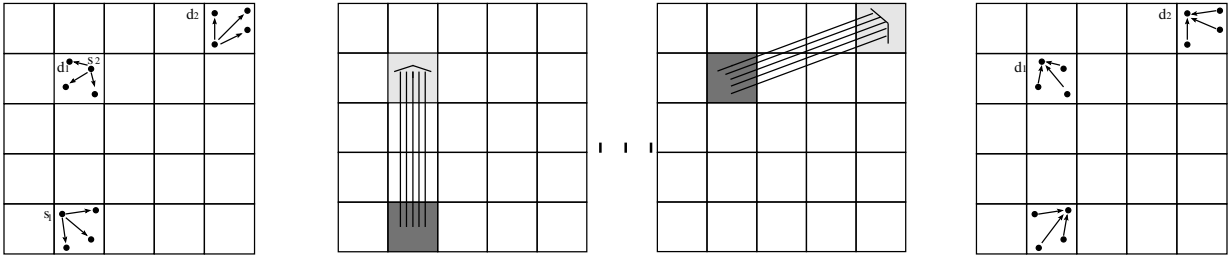}
\end{center}
\caption{The three phase scheme with better scheduling. The figure illustrates the operation in one session.}
\label{fig:delay_ph3}
\end{figure*}

\textbf{Phase 1: Setting Up Transmit Cooperation} Clusters work in
parallel. The chosen source node in each cluster distributes its $M$
bits to the other nodes by using TDMA, which takes $M$ time-slots in total. Note that as opposed to the scheme described in Section \ref{sec:hier1}, there is only one source node operating in each cluster.

\textbf{Phase 2: MIMO Transmissions} Successive MIMO transmissions originated from each cluster are performed, transmitting the bits of the active source node in each cluster to its respective destination cluster. Note that in the current case, there is only one MIMO transmission originated from each cluster, so there are only $n/M$ MIMO transmissions that need to be performed in total. This will require total time $n/M$.

\textbf{Phase 3: Cooperate to Decode} Clusters work in parallel. Each cluster received at most $\log n$ MIMO transmissions in phase 2 by Lemma~\ref{lem}-b, each MIMO transmission intended for
a different destination node in the cluster. The received observations at each node are quantized into $Q$ bits and are to be conveyed to the actual destination nodes. The traffic inside each cluster is at most of exchanging $QM\log n$ bits and can be completed using TDMA in at most $QM\log n$ time slots. (See Figure~\ref{fig:delay_ph3}.)

The operation continues with the next session by choosing a new set of $n/M$ source nodes randomly among the nodes that have not yet accomplished the above steps. Note that all source-destination pairs will accomplish the three steps in a total of $M$ sessions. 

\medbreak
With this rather smoother operation on the network level, we accomplish to serve $n/M$ source-destination pairs in each session, that is transfer $M\times \frac{n}{M}$ bits in total to their destinations in $M+\frac{n}{M}+QM\log n$ time slots yielding aggregate throughput
\begin{equation}\label{th_sc}
\frac{M\times \frac{n}{M}}{M+\frac{n}{M}+QM\log n} 
\end{equation}
which is maximized by the choice $M=\sqrt{n}$ yielding aggregate throughput $T(n)=\frac{1}{2+Q}\frac{\sqrt{n}}{\log n}$. The delay experienced by each bit is now much less compared to the three phase scheme in Section~\ref{sec:hier1}, since it is again dictated by the total time spent in the three phases (denominator of (\ref{th_sc})), which is now less than $D(n)=(2+Q)\sqrt{n}\log n$. 

Note that instead of choosing $M=\sqrt{n}$, which is the optimal choice to maximize the throughput achieved by the scheme, one can choose $M=n^b$ with $0\leq b\leq 1/2$. In this case, we also restrict the number of source-destination pairs to be served in each session to $M$, which can be less than the total number of clusters $n/M$. Indeed, we operate one source node in each of the $M (\leq n/M)$ clusters and simply keep the remaining clusters inactive. The expression for the aggregate throughput becomes
\begin{equation*}
\frac{M\times M}{M+M+QM\log n} 
\end{equation*}
which implies that the scheme achieves aggregate throughput $T(n)= n^b/\log n$ and delay $D(n)= n^b\log n$ for any $0\leq b\leq 1/2$. Note that this throughput-delay trade-off differs only by $\log{n}$ from the trade-off achieved by multi-hop schemes.

\subsection{Better Scheduling for the Hierarchical Cooperation Scheme}\label{sec:bshcs}
In this section, we adopt the scheduling idea of Section~\ref{sec:bstps} to the modified hierarchical scheme presented in Section~\ref{sec:sbs}. However, this modification is not trivial and requires us to consider a generalized version of the network multiple access problem.
\smallbreak

\begin{definition}[The Generalized Network MAC Problem]\label{def:gnma}
Consider the assumptions on the network and channel model given in Section~\ref{sec:result}. Let each of the $n$ nodes in the network be interested in conveying independent information to a subset $A(n)$ of the nodes ($A(n)\leq n$), where the $A(n)$ nodes are chosen randomly among the $n$ nodes in the network. In particular, let us assume that each node in the network has an independent $1$ bit message (or $L$ independent bits, with $L$ constant) to send to each of these $A(n)$ nodes and the quantity of interest is the minimal time $G(n)$ required to accomplish this task. We define this to be the generalized network multiple access problem.
\end{definition}
\smallbreak
The following theorem provides an achievable solution to this problem. We skip the proof of the theorem since it is similar in spirit to the proof of Theorem~\ref{thm1}.
\smallbreak
\begin{theorem}\label{thm2}
For any integer $h>0$, if $A(n) \geq n^{\frac{h}{h+1}}$, then the network multiple access problem can be solved in
$$
G(n)\leq K \, \frac{A(n)}{n} \, n^{\frac{h+1}{h}}\log(n)
$$
time-slots w.h.p., for some constant $K>0$ independent of $n$.
\end{theorem}
\smallbreak
Note that the generalized network multiple access problem contains the network multiple access problem discussed earlier as a special case with $A(n)=n$. Plugging $A(n)=n$ in Theorem~\ref{thm2}, we recover the result of Theorem~\ref{thm1} with an extra $\log n$ factor. Indeed, when the condition $A(n)\geq n^{\frac{h}{h+1}}$ is satisfied with strict inequality in order, the extra $\log n$ factor in Theorem~\ref{thm2} is not needed. However, it is needed to account for the case $A(n)= n^{\frac{h}{h+1}}$, in which case it arises due to part-b of Lemma~\ref{lem}.   
\smallbreak

We are now ready to apply the scheduling idea in Section~\ref{sec:bstps} to the hierarchical cooperation scheme. Consider dividing the network into clusters of $M_1$ nodes and then further divide these clusters into smaller clusters of size $M_2$. Following the scheduling idea in Section~\ref{sec:bstps}, let us organize $M_1/M_2$ sessions and for each session randomly choose one small cluster inside every large cluster. Only the source nodes located in the chosen small clusters and their corresponding destination nodes will be served at each session. As usual, we are operating in three successive phases in each session:

\textbf{Phase 1: Setting Up Transmit Cooperation} The active small clusters operate in parallel. Note that there is a single active cluster of size $M_2$ inside every large cluster of size $M_1$. Let $S$ be the chosen small cluster inside the larger cluster $L$ that will operate in the current session. In this phase, each of the $M_2$ source nodes in $S$ need to distribute their $M_1$ bits among the $M_1$ nodes in the larger cluster $L$, each of the $M_1$ bits goes to a different node. This can be accomplished in two sub-phases:

\begin{itemize}
\item \textbf{Sub-Phase 1: MIMO transmissions} Successive MIMO transmissions are performed between nodes in $S$ and each node in $L$. In each of these MIMO transmissions, say the one between $S$ and a node $d$ in $L$ (located outside of $S$), the $M_2$ nodes in $S$ are simultaneously transmitting the $1$ bit messages they would like to communicate to $d$. The $M_2$ nodes located in the same small cluster with $d$ are acting as a distributed receive antenna array for this MIMO transmission. Since these MIMO transmissions should be repeated for every node in $L$, this sub-phase takes a total of $M_1$ time-slots. See Figure~\ref{fig:sch:subph1}.

\begin{figure}[tb]
\begin{center}
\includegraphics[width=0.3\textwidth]{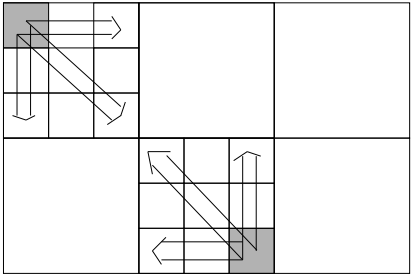}
\end{center}
\caption{The figure illustrates sub-phase 1 of phase 1 of the modified hierarchical scheme with better scheduling. Note that there is only one small cluster distributing bits inside every large cluster.}
\label{fig:sch:subph1}
\end{figure}
\item 
\textbf{Sub-Phase 2: Cooperate to Decode} All small clusters in the network work in parallel. In particular, each small cluster in $L$ has received $M_2$ MIMO transmissions from $S$ in the previous phase, one MIMO transmission for each node in this small cluster. Thus, each node in the small cluster has $M_2$ observations, one from each of the MIMO transmissions and each observation is to be conveyed to a different node in the cluster. Quantizing each observation into $Q$ bits, we get the network multiple access problem defined in Section~\ref{sec:sbs} in a network of size $M_2$, and by Theorem~\ref{thm1} this problem can be handled in $QM_2^{\frac{h_1+1}{h_1}}$ time-slots for any integer $h_1>0$. 
\end{itemize}

\textbf{Phase 2: MIMO Transmissions} At the end of the first phase, all source nodes in the active small clusters have distributed their $M_1$ bits among the nodes in the larger cluster. Now, successive long-distance $M_1\times M_1$ MIMO transmissions between large clusters are performed. During each MIMO transmission, the $M_1$ bits of a particular source node in the active small cluster are transferred to the destination cluster where its destination node is located. The number of MIMO transmissions to be performed in this phase is equal to the total number of source nodes active in this session. Hence the total phase can be completed in $\frac{n}{M_1}\times M_2$ time-slots.

\textbf{Phase 3: Cooperate to Decode} By part-a of Lemma~\ref{lem}, there are order $M_2$ destination nodes located in each of the large clusters. Thus, each large cluster has received $M_2$ MIMO transmissions in the previous phase, and the quantized MIMO observations spread over the $M_1$ nodes of the large cluster should be collected at the corresponding $M_2$ destination nodes. This is the generalized network multiple access problem of size $M_1$ with $A(M_1)=M_2$. By Theorem~\ref{thm2}, it can be solved in $Q\frac{M_2}{M_1}\times M_1^{\frac{h_2+1}{h_2}}\log M_1$ time-slots for any integer $h_2> 0$ provided that $A(M_1)\geq M_1^{\frac{h_2}{h_2+1}}$.

Gathering everything together, at every session of this modified hierarchical cooperation scheme, we deliver $M_1\times M_2\times\frac{n}{M_1}$ bits to their destinations in a total of $$\left(M_1+QM_2^{\frac{h_1+1}{h_1}}\right)+\frac{n}{M_1}\times M_2+Q\frac{M_2}{M_1}\times M_1^{\frac{h_2+1}{h_2}}\log M_1$$ time-slots. The aggregate throughput is given by
$$
\frac{\frac{n}{M_1}\times M_2\times M_1}{M_1+QM_2^{\frac{h_1+1}{h_1}}+\frac{n}{M_1}\times M_2+Q\frac{M_2}{M_1}\times M_1^{\frac{h_2+1}{h_2}}\log M_1}
$$
which is maximized by the choice $h=h_2=h_1+1$, $M_1=n^{\frac{h}{h+1}}$ and $M_2=M_1^{\frac{h-1}{h}}$, yielding aggregate throughput $T(n)=\frac{n^{\frac{h}{h+1}}}{\log n}$ and delay $D(n)=n^{\frac{h}{h+1}}\log n$. Note that these choices for $M_1$ and $M_2$ satisfy the constraint $A(M_1) = M_2 \geq M_1^{\frac{h_2}{h_2+1}}$. 
\smallbreak
Note that at this point, we have proven that all points on the throughput-delay scaling curve $(T(n),D(n))=(n^{\frac{h}{h+1}}/\log n,\, n^{\frac{h}{h+1}}\log n)$ with $h$ being a positive integer are achievable. In order to show that all points on the line $(T(n),D(n))=(n^{b}/\log n,\, n^{b}\log n)$ with $0\leq b< 1$ are achievable, we can choose $M_1=n^b$ with $0\leq b\leq \frac{h}{h+1}$ in the above discussion, while maintaining the relationships $M_2=M_1^{\frac{h-1}{h}}$ and $h=h_2=h_1+1$. Extending the argument at the end of Section~\ref{sec:bstps}, we also restrict the number of small clusters to be served in each session to $M_1^{1/h}$ which can now be less than the total number of large clusters $n/M_1\,(\,\geq M_1^{1/h})$. Indeed, we operate one small cluster in each of the $M_1^{1/h}$ large clusters and simply keep the remaining large clusters inactive. The expression for the aggregate throughput becomes
$$
\frac{M_1^{\frac{1}{h}}\times M_2\times M_1}{M_1+QM_2^{\frac{h_1+1}{h_1}}+M_1^{\frac{1}{h}}\times M_2+Q\frac{M_2}{M_1}\times M_1^{\frac{h_2+1}{h_2}}\log M_1}
$$
which shows that we can achieve aggregate throughput $T(n)= M_1/\log M_1$ and delay $D(n)= M_1\log M_1$. Recalling that $M_1=n^b$, we get the points on the throughput-delay scaling curve $(T(n),D(n))=(n^b/\log n,\, n^b \log n)$ for any $0\leq b\leq \frac{h}{h+1}$ and $h>0$.
This concludes the proof of the main result of this paper.\hfill $\square$

\section{Conclusion}
The present work shows that hierarchical cooperation not only can lead to higher throughput in ad hoc networks, but also to reasonable end-to-end delay, given that some extra care is taken in setting up cooperation at the lower levels and scheduling communications. Meanwhile, we have discussed the network multiple-access problem in the present paper, which is of interest in its own right.

\begin{biographynophoto}
Ayfer {\"O}zg{\"u}r received B.Sc. degrees in electrical engineering and physics from Middle East Technical University, Turkey, in 2001 and M.Sc. degree in electrical engineering from the same university in 2004. She is now a Ph.D. student at the Laboratory of Information Theory, Swiss Federal Institute of Technology-Lausanne. Her research interests include wireless communications and information theory.
\end{biographynophoto}

\begin{biographynophoto}
Olivier L\'ev\^eque was born in Switzerland in 1971. He received the 
physics diploma from EPFL in 1995 and completed his Ph.D. in mathematics 
at EPFL in 2001. Since then, he has been with the Laboratory of 
Information Theory at EPFL. He spent the academical year 2005-2006 at 
the Electrical Engineering Department of Stanford University, where he 
was appointed as lecturer. His research interests include stochastic 
analysis, random matrices, wireless communications and information theory.
\end{biographynophoto}

\end{document}

%% file: delay-thrput_tradeoff_3.pstex_t
\begin{picture}(0,0)%
\includegraphics{delay-thrput_tradeoff_3.pstex}%
\end{picture}%
\setlength{\unitlength}{4144sp}%
\begingroup\makeatletter\ifx\SetFigFont\undefined%
\gdef\SetFigFont#1#2#3#4#5{%
  \reset@font\fontsize{#1}{#2pt}%
  \fontfamily{#3}\fontseries{#4}\fontshape{#5}%
  \selectfont}%
\fi\endgroup%
\begin{picture}(3625,2439)(-224,-1726)
\put(2206,479){\makebox(0,0)[lb]{\smash{{\SetFigFont{9}{10.8}{\familydefault}{\mddefault}{\updefault}Mobility Schemes}}}}
\put(316,-1681){\makebox(0,0)[lb]{\smash{{\SetFigFont{9}{10.8}{\familydefault}{\mddefault}{\updefault}$1$}}}}
\put( 91,-1501){\makebox(0,0)[lb]{\smash{{\SetFigFont{9}{10.8}{\familydefault}{\mddefault}{\updefault}$1$}}}}
\put(541,-376){\makebox(0,0)[lb]{\smash{{\SetFigFont{9}{10.8}{\familydefault}{\mddefault}{\updefault}Multi-hop}}}}
\put(1216,-1681){\makebox(0,0)[lb]{\smash{{\SetFigFont{9}{10.8}{\familydefault}{\mddefault}{\updefault}$\sqrt{n}$}}}}
\put(2971,-1636){\makebox(0,0)[lb]{\smash{{\SetFigFont{14}{16.8}{\familydefault}{\mddefault}{\updefault}T(n)}}}}
\put(-44,569){\makebox(0,0)[lb]{\smash{{\SetFigFont{12}{14.4}{\familydefault}{\mddefault}{\updefault}D(n)}}}}
\put(2071,-1681){\makebox(0,0)[lb]{\smash{{\SetFigFont{9}{10.8}{\familydefault}{\mddefault}{\updefault}$n$}}}}
\put(1261,-961){\makebox(0,0)[lb]{\smash{{\SetFigFont{9}{10.8}{\familydefault}{\mddefault}{\updefault}Cooperation}}}}
\put(1261,-826){\makebox(0,0)[lb]{\smash{{\SetFigFont{9}{10.8}{\familydefault}{\mddefault}{\updefault}Hierarchical}}}}
\put(-134,-691){\makebox(0,0)[lb]{\smash{{\SetFigFont{9}{10.8}{\familydefault}{\mddefault}{\updefault}$\sqrt{n}$}}}}
\put(1711,-1681){\makebox(0,0)[lb]{\smash{{\SetFigFont{9}{10.8}{\familydefault}{\mddefault}{\updefault}$\frac{n}{\log n}$}}}}
\put(901,-1681){\makebox(0,0)[lb]{\smash{{\SetFigFont{9}{10.8}{\familydefault}{\mddefault}{\updefault}$\frac{\sqrt{n}}{\log n}$}}}}
\put(-224,-511){\makebox(0,0)[lb]{\smash{{\SetFigFont{9}{10.8}{\familydefault}{\mddefault}{\updefault}$\sqrt{n}\log n$}}}}
\put(-134,209){\makebox(0,0)[lb]{\smash{{\SetFigFont{9}{10.8}{\familydefault}{\mddefault}{\updefault}$n\log n$}}}}
\end{picture}%

%% file: phase1_3d_ver2.pstex_t
\begin{picture}(0,0)%
\includegraphics{phase1_3d_ver2.pstex}%
\end{picture}%
\setlength{\unitlength}{4144sp}%
\begingroup\makeatletter\ifx\SetFigFont\undefined%
\gdef\SetFigFont#1#2#3#4#5{%
  \reset@font\fontsize{#1}{#2pt}%
  \fontfamily{#3}\fontseries{#4}\fontshape{#5}%
  \selectfont}%
\fi\endgroup%
\begin{picture}(8034,3186)(-4781,-2407)
\put(-3104,-1952){\makebox(0,0)[lb]{\smash{{\SetFigFont{8}{9.6}{\familydefault}{\mddefault}{\updefault}{PHASE 1}%
}}}}
\put(-1369,-1952){\makebox(0,0)[lb]{\smash{{\SetFigFont{8}{9.6}{\familydefault}{\mddefault}{\updefault}{PHASE 2}%
}}}}
\put(310,-1952){\makebox(0,0)[lb]{\smash{{\SetFigFont{8}{9.6}{\familydefault}{\mddefault}{\updefault}{PHASE 3}%
}}}}
\put(-3601,-2143){\makebox(0,0)[lb]{\smash{{\SetFigFont{6}{7.2}{\familydefault}{\mddefault}{\updefault}{PHASE 1}%
}}}}
\put(-3047,-2143){\makebox(0,0)[lb]{\smash{{\SetFigFont{6}{7.2}{\familydefault}{\mddefault}{\updefault}{PHASE 2}%
}}}}
\put(-2493,-2143){\makebox(0,0)[lb]{\smash{{\SetFigFont{6}{7.2}{\familydefault}{\mddefault}{\updefault}{PHASE 3}%
}}}}
\put(-277,-2143){\makebox(0,0)[lb]{\smash{{\SetFigFont{6}{7.2}{\familydefault}{\mddefault}{\updefault}{PHASE 1}%
}}}}
\put(276,-2143){\makebox(0,0)[lb]{\smash{{\SetFigFont{6}{7.2}{\familydefault}{\mddefault}{\updefault}{PHASE 2}%
}}}}
\put(831,-2143){\makebox(0,0)[lb]{\smash{{\SetFigFont{6}{7.2}{\familydefault}{\mddefault}{\updefault}{PHASE 3}%
}}}}
\put(-2294,614){\makebox(0,0)[lb]{\smash{{\SetFigFont{8}{9.6}{\familydefault}{\mddefault}{\updefault}{PHASE 1}%
}}}}
\put(-359,704){\makebox(0,0)[lb]{\smash{{\SetFigFont{8}{9.6}{\familydefault}{\mddefault}{\updefault}{PHASE 2}%
}}}}
\put(1576,569){\makebox(0,0)[lb]{\smash{{\SetFigFont{8}{9.6}{\familydefault}{\mddefault}{\updefault}{PHASE 3}%
}}}}
\put(182,-689){\makebox(0,0)[lb]{\smash{{\SetFigFont{5}{6.0}{\familydefault}{\mddefault}{\updefault}{PHASE 1}%
}}}}
\put(2252,-713){\makebox(0,0)[lb]{\smash{{\SetFigFont{5}{6.0}{\familydefault}{\mddefault}{\updefault}{PHASE 3}%
}}}}
\put(991,-646){\makebox(0,0)[lb]{\smash{{\SetFigFont{5}{6.0}{\familydefault}{\mddefault}{\updefault}{PHASE 2}%
}}}}
\put(-1843,-758){\makebox(0,0)[lb]{\smash{{\SetFigFont{5}{6.0}{\familydefault}{\mddefault}{\updefault}{PHASE 3}%
}}}}
\put(-3779,-781){\makebox(0,0)[lb]{\smash{{\SetFigFont{5}{6.0}{\familydefault}{\mddefault}{\updefault}{PHASE 1}%
}}}}
\put(-3014,-691){\makebox(0,0)[lb]{\smash{{\SetFigFont{5}{6.0}{\familydefault}{\mddefault}{\updefault}{PHASE 2}%
}}}}
\end{picture}%

%% file: delay_arxiv_0225.bbl
\begin{thebibliography}{1}

\bibitem{GK00} P.~Gupta and P.~R.~Kumar, \emph{The Capacity of Wireless Networks}, IEEE Trans. on Information Theory 42 (2), pp.~388--404, March 2000.

\bibitem{GMPS06-I} A.~El~Gamal, J.~Mammen, B.~Prabhakar, D.~Shah, \emph{Optimal Throughput-Delay Scaling in Wireless Networks-Part~I: The Fluid Model},  IEEE Trans. on Information Theory 52(6), pp.~2568-2592, 2006.

\bibitem{GMPS06-II} A.~El~Gamal, J.~Mammen, B.~Prabhakar, D.~Shah, \emph{Optimal Throughput-Delay Scaling in Wireless Networks-Part~II: Constant-Size Packets}, IEEE Trans. on Information Theory 52(11),  pp.~5111-5116, 2006.

\bibitem{OLT07} A.~{\"O}zg{\"u}r, O.~L{\'e}v{\^e}que, D.~Tse, \emph{Hierarchical cooperation achieves optimal capacity scaling in ad hoc networks}, IEEE Transactions on Information Theory,
Vol. 53 (10), pp. 3549-3572, October 2007.

\bibitem{GT02} M. Grossglauser and D. Tse, {\em Mobility Increases the
Capacity of Adhoc Wireless Networks},  IEEE/ACM Transactions on Networking 10(4), pp.~477-486, 2002.

\bibitem{NM05} M. J. Neely and E. Modiano, {\em Capacity and Delay Tradeoffs for Ad-Hoc Mobile Networks}, IEEE Transactions on Information Theory, Vol. 51 (6), pp. 1917-1937, June 2005. 

\bibitem{SMS06} G. Sharma, R. R. Mazumdar and N. B. Shroff, {\em Delay and Capacity Trade-offs in Mobile Ad Hoc Networks: A Global Perspective}, 2006 IEEE INFOCOM Conference, Barcelona, Spain, April 2006.

\bibitem{NMC07} M. Neely, E. Modiano and Y. S. Cheng, {\em Logarithmic Delay for $N\times N$ Packet Switches Under the Crossbar Constraint}, IEEE/ACM Trans. on Networking, Vol. 15 (3), June 2007.

\bibitem{Gal91} Gallager R. G., {\em Information Theory and Reliable Communication}, Wiley \& Sons Inc., 1968.

\bibitem{Tel99} E.~Telatar, \emph{Capacity of Multi-Antenna Gaussian Channels"} European Trans.~on Telecommunications, ETT, vol.10 (6), pp. 585-596, Nov. 1999.
\end{thebibliography}
